\documentclass[aps,superscriptaddress,floatfix,preprint]{revtex4}
\usepackage{epsfig,amssymb,amsmath}
\begin{document}
\title{Possible Connection between the Optimal Path and Flow\\ in Percolation Clusters}

\author{Eduardo L\'{o}pez}
\affiliation{Center for Polymer Studies, Boston University, Boston, MA 02215, USA}
\author{Sergey V. Buldyrev}
\affiliation{Center for Polymer Studies, Boston University, Boston, MA 02215, USA}
\affiliation{Department of Physics, Yeshiva University, 500 W 185th Street, New York, NY 10033}
\author{Lidia A. Braunstein}
\affiliation{Center for Polymer Studies, Boston University, Boston, MA 02215, USA}
\affiliation{Departamento de F\'{i}sica, Facultad de Ciencias Exactas y
Naturales, Universidad Nacional de Mar del Plata, Funes 3350, 7600 Mar del Plata, Argentina}
\author{Shlomo Havlin}
\affiliation{Minerva Center \& Department of Physics, Bar-Ilan University, Ramat Gan, Israel}
\author{H. Eugene Stanley}
\affiliation{Center for Polymer Studies, Boston University, Boston, MA 02215, USA}

\date{lbbhs.tex ~~~ \today}

\begin{abstract}

  We study the behavior of the optimal path between two sites separated
  by a distance $r$ on a $d$-dimensional lattice of linear size $L$ with
  weight assigned to each site. We focus on the strong disorder limit,
  i.e., when the weight of a single site dominates the sum of the
  weights along each path.  We calculate the probability distribution
  $P(\ell_{\rm opt}|r,L)$ of the optimal path length $\ell_{\rm opt}$,
  and find for $r\ll L$ a power law decay with $\ell_{\rm opt}$,
  characterized by exponent $g_{\rm opt}$.  We determine the scaling
  form of $P(\ell_{\rm opt}|r,L)$ in two- and three-dimensional
  lattices.  To test the conjecture that the optimal paths in strong
  disorder and flow in percolation clusters belong to the same
  universality class, we study the tracer path length $\ell_{\rm tr}$ of
  tracers inside percolation through their probability distribution
  $P(\ell_{\rm tr}|r,L)$.  We find that, because the optimal path is not
  constrained to belong to a percolation cluster, the two problems are
  different.  However, by constraining the optimal paths to remain
  inside the percolation clusters in analogy to tracers in percolation,
  the two problems exhibit similar scaling properties.

\end{abstract}
\maketitle

\section{Introduction}

Flow in porous media, due to its ubiquitous nature, has 
received a great deal of attention in recent decades 
\cite{ben-Avraham,Shlomo-lmin,Bunde,Andrade,Lopez}.
Interest has been driven by multiple real-world problems such as oil extraction
\cite{Andrade,Lopez} and ground water polution \cite{Koplik-Redner-Wilkinson,Bacri}.
Porous media is typically simulated through percolation systems \cite{Stauffer}, and flow 
through a variety of models ranging from the idealized ant in a labyrinth \cite{deGennes}
to the very practical ones like convective tracer flow \cite{Sahimi_pla,Lopez,Koplik}.

In the study of flow in percolation clusters, it has been hypohesize that
transport properties may be related to some geometric property of the cluster. 
However, despite considerable effort \cite{ben-Avraham} such a geometric
property relation has not been found \cite{Shlomo-lmin}. 
Here, we argue that, for the particular case of convective tracer flow
in percolation clusters, a connection exists between transport properties
and the static properties of a model of optimal paths in disordered lattices,
described in detail below. This relation allows for a mapping between
the two problems that opens new approaches for their study.

The first hint of a possible connection between flow in porous media
and the optimal path problem arouse in the work of Lee et al. \cite{Lee}.
This reference deals with a simplified model of flow in porous media related to
secondary oil extraction, in which an invading fluid (water, steam, etc.) injected at $A$
pushes oil out through extraction well $B$. This method is used when the oil reservoir 
does not have enough pressure to be exploited without this added driving pressure.
The system is considered in the following way: the medium is modeled by a percolacion
cluster at the critical threshold density, and a steady state laminar flow of an incompresible 
fluid is established between $A$ and $B$. Then, tracers which mimick the flow of the
driving fluid are injected at $A$ and travel purely by convection to $B$ (Sec. \ref{comparison}). 
The tracer length ($\ell_{\rm tr}$) probability density function (PDF), $P(\ell_{\rm tr})$,
was measured and it was found that 
the most probable traveling length $\ell^*_{\rm tr}$, defined as the
value of $\ell_{\rm tr}$ for which $P(\ell_{\rm tr})$ is maximum, 
scales with the distance $r$ between the injection and extraction sites $A$ and $B$ as
\begin{equation}
\ell^*_{\rm tr}\sim r^{d_{\rm tr}} \qquad [d_{\rm tr}=1.21\pm 0.02 ~~
  \mbox{for} ~~ d=2].
\label{ltr_r}
\end{equation}
The remarkable feature of Eq.~(\ref{ltr_r}) is that, for $d=2$, the exponent 
$d_{\rm tr}\approx d_{\rm opt}$, where $d_{\rm opt}$ is the exponent for the optimal 
path length in the strong disorder limit as defined by Cieplak et al. 
\cite{Cieplak,Porto}(see below).
In this study, we propose that transport in percolation can be directly
related to the optimal path in strong disorder, and we support our
proposal with extensive numerical simulations for $d$-dimensional
lattices with $d=2$ and 3.

The optimal path problem formulated by Cieplak in Ref.~\cite{Cieplak} is that of finding the
path of lowest cost to go from one end of a $d$-dimensional lattice to the other
end when to
each site (or bond) $i$, we associate a weight $\epsilon_i=e^{ax_i}$, with
$x_i\in [0,1)$. This is equivalent to choosing $\epsilon_i$ from the
distribution
\begin{equation}
W(\epsilon)=\frac{1}{a\epsilon}\qquad \epsilon\in [1,\exp a]. 
\label{We}
\end{equation}
The energy of any path of length $\ell$ on the lattice is given by the sum
\begin{equation}
E\equiv\sum_j^\ell \epsilon_j,
\label{energy}
\end{equation}
where $j$ is an index running over the sites of the path.
The limit $a\rightarrow\infty$ is known as the strong disorder limit.
The optimal path of length $\ell_{\rm opt}$ is the path for which $E$ is
minimal with respect to all other paths. The optimal path length scales
with $r$ as \cite{Cieplak}
\begin{equation}
\ell_{\rm opt}\sim r^{d_{\rm opt}}\qquad [d_{\rm opt}=1.22\pm 0.01].
\label{lopt_r}
\end{equation}
The first reference that we have found to this problem, although 
formulated in a different context is that of Ambegaokar \cite{Ambegaokar}, 
where the metal-insulator transition is considered by a percolation
model in which sites represent possible electron states which can be 
reached only by hopping through quantum tunneling, and the
hopping rates are exponential, motivating Eq.~(\ref{We}).
Another context in which these weights have been observed is that
of magnetoresistance in thin Ni films \cite{Strelniker}.

Our work can be equated to the following question: 
is flow in percolation clusters, a dynamical process, connected to the 
optimal path length in strong disorder, a static property, 
as suggested by the similarity of Eqs.~(\ref{ltr_r})
and (\ref{lopt_r})? To this end, we study the 
PDF $P(\ell_{\rm opt}|r,L)$ for the optimal path to have a
length $\ell_{\rm opt}$, given a system size $L$, and an Euclidean
distance $r$ between the starting and ending sites $A$ and $B$ of the
path. We will compare this PDF with $P(\ell_{\rm tr}|r,L)$, the
PDF that convective tracer paths have a length $\ell_{\rm tr}$ in a
percolation system of size $L$ at criticality, where the starting and ending sites
are at a distance $r$.

There are some indications that this connection is indeed present,
given that other relations between percolation and strong disorder optimal paths
have been reported. For instance, for a lattice with disorder given by
Eq.~(\ref{We}) in the strong disorder limit, the most probable largest
weight of the site used by the optimal path
is $e^{ap_c}$, where $p_c$ is the percolation threshold \cite{Porto}.
Also, Wu et al.~\cite{Wu} recently determined through
the study of tracer flow on a lattice with disorder, that
the strong disorder limit has a length scale that scales as
$a^{\nu}$, where $\nu$ is the connectedness exponent of percolation \cite{Bunde},
and hence a system is in the strong disorder limit only when $a^{\nu}>L$. In this paper,
we study the ultrametric limit $a\rightarrow\infty$.

In Sec.~\ref{Pl_opt} we present results for the distribution
$P(\ell_{\rm opt}|r,L)$. In Sec.~\ref{comparison}, results for
$P(\ell_{\rm tr}|r,L)$ are presented and compared with $P(\ell_{\rm
opt}|r,L)$.  We then discuss the results in Sec.~\ref{discussion}.

\section{Optimal Path Distribution}
\label{Pl_opt}

To study $P(\ell_{\rm opt}|r,L)$ we use the ``bombing algorithm''
proposed in Ref.~\cite{Cieplak}. The optimal path length $\ell_{\rm opt}$
between sites $A$ and $B$ is found by eliminating (bombing)
sites of the lattice in decreasing order of weight, but leaving those
sites necessary to keep $A$ and $B$ connected. When all sites that
do not disconnect $A$ and $B$ are eliminated, only the sites of the
optimal path remain \cite{Braunstein,Buldyrev-algorithm,Text1}.

\begin{figure}
\epsfig{file=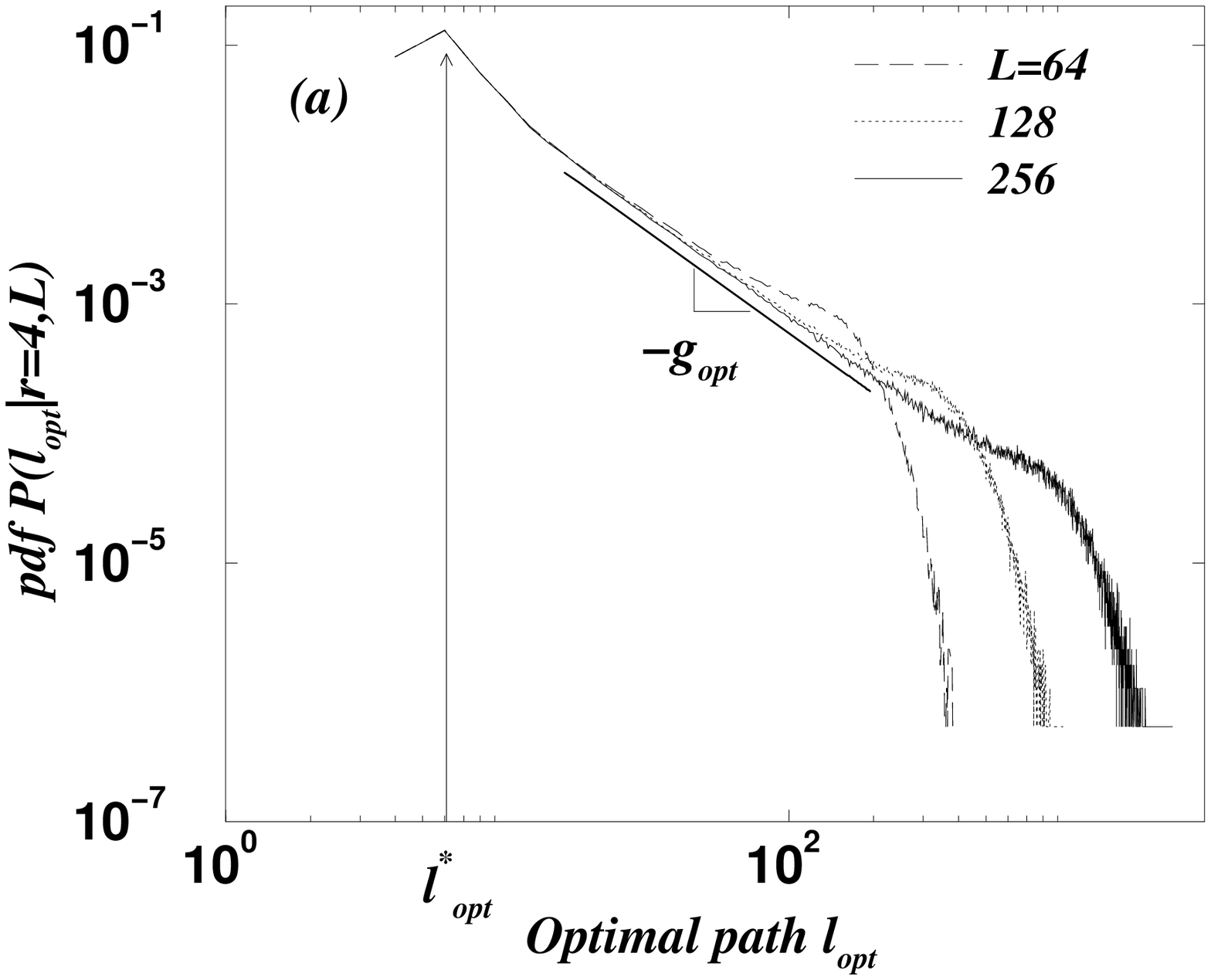,height=7.5cm,width=8.1cm}
\epsfig{file=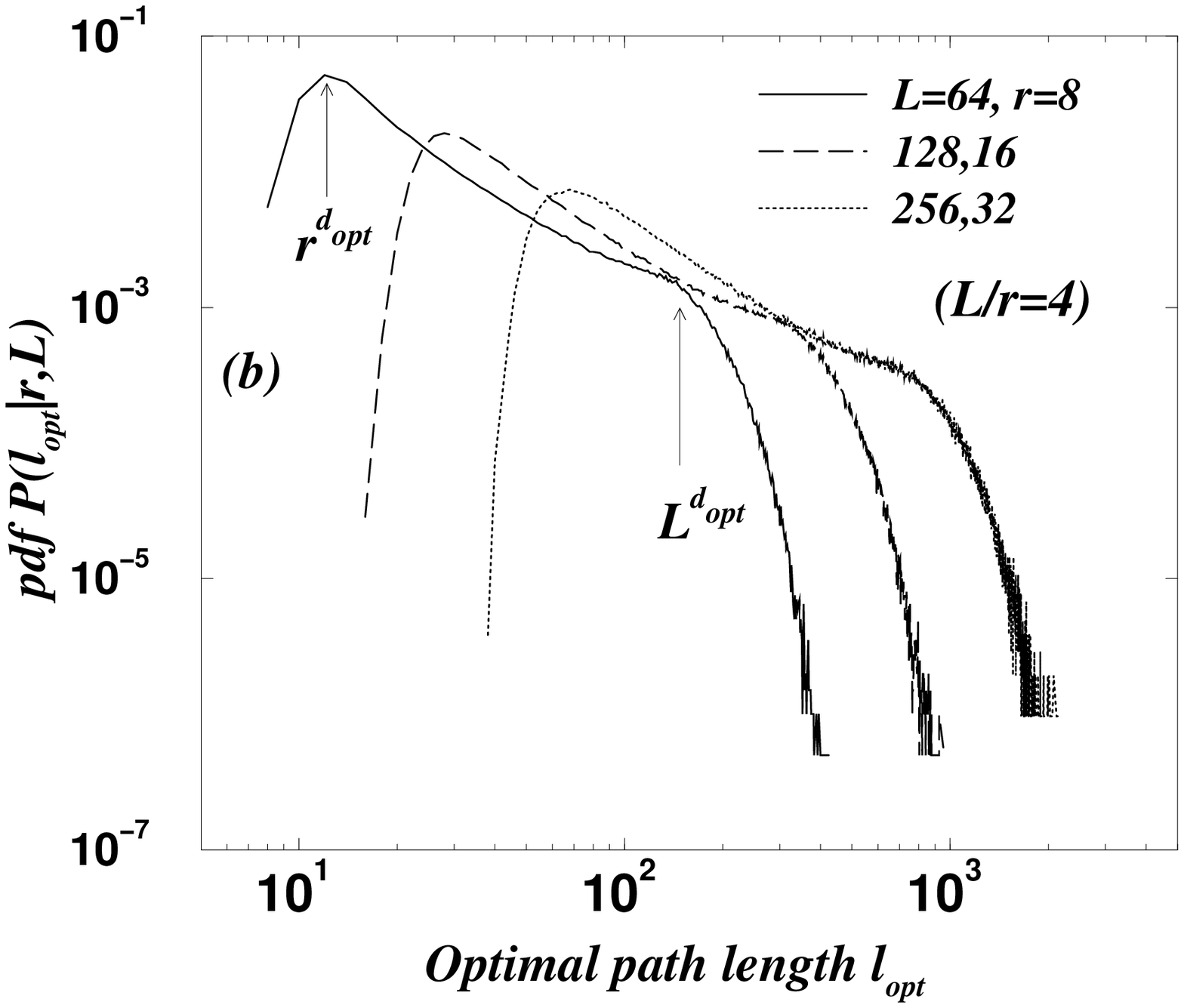,height=7.5cm,width=8.1cm}
\epsfig{file=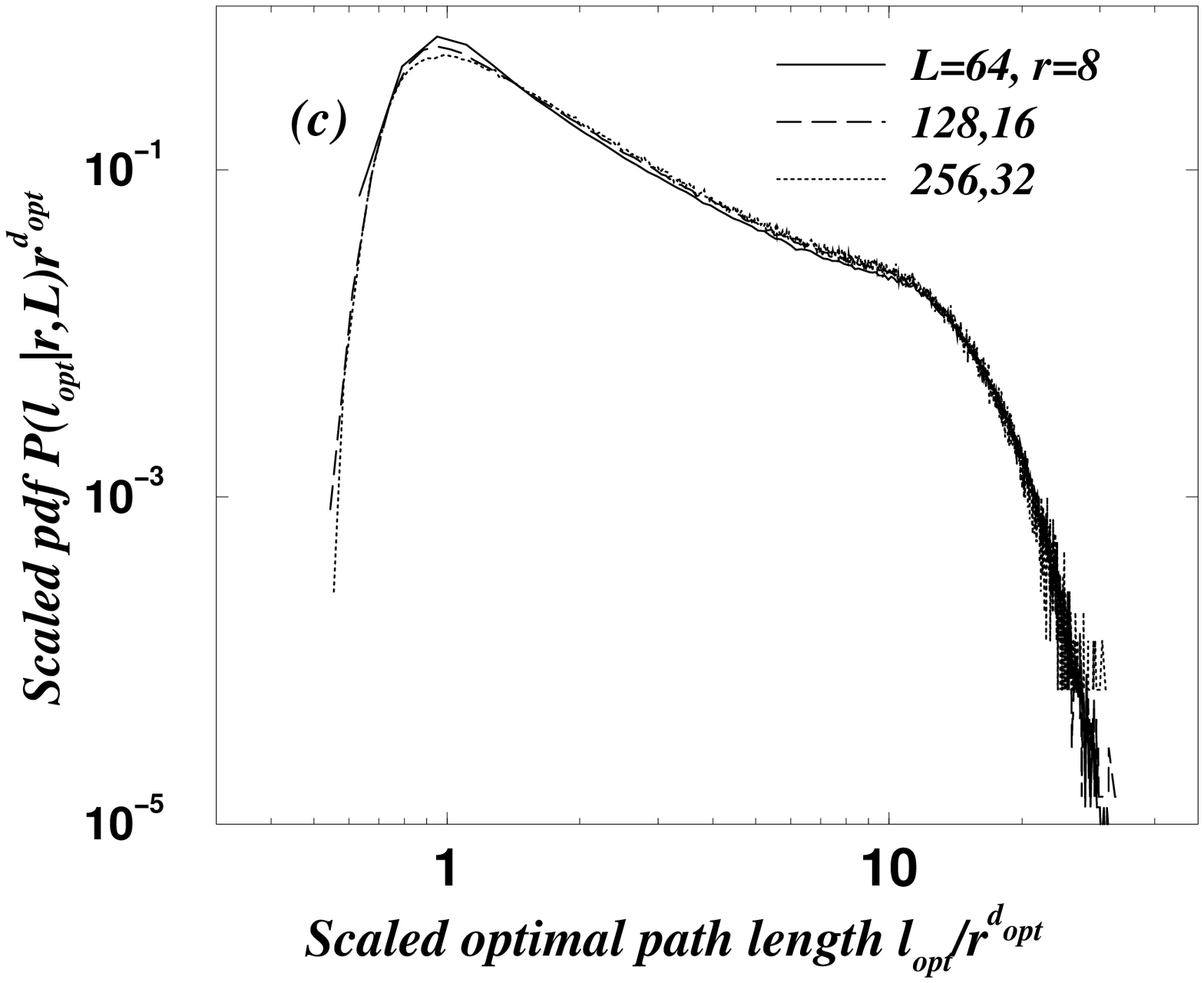,height=7.5cm,width=8.1cm}
\caption{(a) Distribution $P(\ell_{\rm opt}|r,L)$ for $r=4$ and system
  sizes $L=64, 128$ and 256. As $L$ increases, the power law region with
  exponent $g_{\rm opt}$ becomes better defined, and the upper cutoff
  increases.  (b) Probability distribution $P(\ell_{\rm opt}|r,L)$ for
  $(r=8, L=64)$ (solid line), $(r=16, L=128)$ (dashed line), and $(r=32,
  L=256)$ (dotted line) for two-dimensional systems. The ratio between
  $L$ and $r$ is kept fixed for these curves.  (c) Scaled distribution
  $r^{d_{\rm opt}}P(\ell_{\rm opt}|r,L)$ vs. scaled optimal path length
  $\ell_{\rm opt}/r^{d_{\rm opt}}$ for the curves in (b). 
  The collapse has been achieved using
  the exponent $d_{\rm opt}$ reported for $\bar{\ell}_{\rm opt}$, which
  is also valid for the most probable length $\ell^*_{\rm opt}$ as
  evidenced in the plot.  }
\label{Pl_l_s4}
\end{figure}

In Fig.~\ref{Pl_l_s4} we present $P(\ell_{\rm opt}|r,L)$ for $r\ll L$
for a square lattice of sites. Four distinct features appear:

\begin{itemize}

\item The most probable optimal path length $\ell_{\rm opt}^*$ scales
  with $r$ as
\begin{equation}
\ell_{\rm opt}^*\sim r^{d_{\rm opt}}.
\end{equation}
The values of $d_{\rm opt}$ have been reported for several lattice
dimensions $d$, and also have been shown to be universal
\cite{Buldyrev}.  Here, we rescale $P(\ell_{\rm opt}|r,L)$ with the
exponent $d_{\rm opt}$, calculated elsewhere for the average
optimal path length $\bar{\ell}_{\rm opt}$, but, as Figs.~\ref{Pl_l_s4}a
and \ref{Pl_l_s4}b show, $d_{\rm opt}$ also produces the correct scaling
for $\ell_{\rm opt}^*$.  Our results for $d_{\rm opt}$ are reported in
Table \ref{table1} for $d=2$ and 3.

\item A lower cutoff [Fig.~\ref{Pl_l_s4}(b)] which, in analogy with the
  distribution of minimal paths in percolation
  \cite{Havlin-ben-Avraham,Dokh,Andrade}, is expected to be a stretched
  exponential function $f_1$ of the form
\begin{equation}
f_1(x)=\exp(-\alpha x^{-\phi_{\rm opt}}) \qquad \left[x\equiv\frac{\ell_{\rm
      opt}}{r^{d_{\rm opt}}}\right],
\label{f_1}
\end{equation}
where $\alpha$ is a lattice-dependent constant, and $\phi_{\rm opt}$ is
a universal exponent satisfying \cite{Buldyrev,Havlin-ben-Avraham}
\begin{equation}
\phi_{\rm opt}=\frac{1}{d_{\rm opt}-1}.
\label{phi_formula}
\end{equation}

\item An upper cutoff due to the effect of the finite lattice size $L$.
  A stretched exponential behavior is also expected to describe this
  region \cite{Buldyrev-algorithm,Dokh,Andrade}, through a function $f_2$
  of the form
\begin{equation}
f_2(y)=\exp(-\beta y^{\psi_{\rm opt}}) \qquad \left[y\equiv\frac{\ell_{\rm
      opt}}{L^{d_{\rm opt}}}\right],
\label{f_2}
\end{equation}
where $\beta$ is a lattice-dependent constant, and $\psi_{\rm opt}$ has
universal properties \cite{Buldyrev-algorithm,Dokh,Andrade}.

\item A power-law region described by
\begin{equation}
P(\ell_{\rm opt})\sim \ell_{\rm opt}^{-g_{\rm opt}} \qquad
[r^{d_{\rm opt}}< \ell_{\rm opt}\le L^{d_{\rm opt}}].
\end{equation}
\end{itemize}

The above considerations lead us to postulate for $P(\ell_{\rm
opt}|r,L)$ a full scaling {\it Ansatz\/}
\cite{Buldyrev,Havlin-ben-Avraham,Dokh,Andrade}
\begin{equation}
P(\ell_{\rm opt}|r,L)\sim {\frac{1}{r^{d_{\rm opt}}}}
\left(\frac{\ell_{\rm opt}}{r^{d_{\rm opt}}}\right)^{-g_{\rm opt}}
f_1\left(\frac{\ell_{\rm opt}}{r^{d_{\rm opt}}}\right)
f_2\left(\frac{\ell_{\rm opt}}{L^{d_{\rm opt}}}\right),
\label{ansatz}
\end{equation}
where the prefactor $1/r^{d_{\rm opt}}$ is necessary for normalization.
We have tested this {\it Ansatz\/} for $d=2$,~3 and found it to be
consistent with our earlier simulations \cite{Buldyrev}.

An interesting feature of $P(\ell_{\rm opt}|r,L)$ is that, as $d$
increases, $g_{\rm opt}$ decreases. In other words, the longer optimal
paths at larger dimensions have a larger probability (see
Table~\ref{table1}).  Additionally, since $g_{\rm opt}<2$ for all $d$,
$\bar{\ell}_{\rm opt}$ and all higher moments diverge as
$L\rightarrow\infty$.

\begin{figure}
\epsfig{file=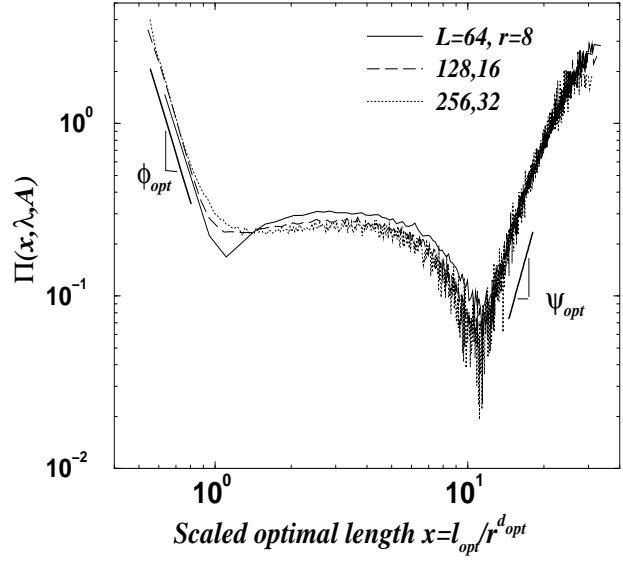,height=7.5cm,width=8.1cm}
\caption{The scaling function $\Pi(x,\lambda,A)$ for $A=0.1$ vs. the
  scaled optimal path length $x\equiv\ell_{\rm opt}/r^{d_{\rm opt}}$ for
  system sizes $(r=8,L=64)$, $(r=16,L=128)$, and $(r=32,L=256)$.  The two
  straight lines serve as guides to the eye for the data that determine
  the exponents $\phi_{\rm opt}$ and $\psi_{\rm opt}$.  }
\label{Pi}
\end{figure}

To calculate the exponents $\phi_{\rm opt}$ and $\psi_{\rm opt}$ of
Eqs.~(\ref{f_1}) and (\ref{f_2}), we introduce the function
\begin{equation}
\Pi\left(\frac{\ell_{\rm opt}}{r^{d_{\rm opt}}},\lambda,A\right)\equiv 
\ln\left[\frac{A}{P(\ell_{\rm opt}|r,L)r^{d_{\rm opt}}
\left(\frac{\ell_{\rm opt}}{r^{d_{\rm opt}}}\right)^{g_{\rm opt}}}\right] 
\end{equation}
which, upon using Eqs.~(\ref{f_1}), (\ref{f_2}) and (\ref{ansatz}) yields
\begin{eqnarray}
  \Pi\left(\frac{\ell_{\rm opt}}{r^{d_{\rm opt}}},\lambda,A\right)&\sim&
  \ln\left[\frac{A}{f_1\left(\frac{\ell_{\rm opt}}{r^{d_{\rm opt}}}\right)
      f_2\left(\frac{\ell_{\rm opt}}{r^{d_{\rm opt}}}\lambda^{-d_{\rm
            opt}}\right)}\right]\nonumber\\
  &\sim&\ln A+\alpha\left(\frac{\ell_{\rm opt}}{ 
      r^{d_{\rm opt}}}\right)^{-\phi_{\rm opt}}
  +\beta\left(\frac{\ell_{\rm opt}}{r^{d_{\rm opt}}}\lambda^{-d_{\rm 
        opt}}\right)^{\psi_{\rm opt}}.
\label{Pi_function}
\end{eqnarray}
We have made use of $\lambda\equiv L/r$ in the argument of
the function $f_2$ so that $f_2(\ell_{\rm opt}/L^{d_{\rm
    opt}})=f_2(\lambda^{-d_{\rm opt}}\ell_{\rm opt}/r^{d_{\rm opt}})$.
The constant $A$ is an auxiliary parameter chosen to make the minimum
value of $\Pi$ slightly larger than unity.  Defining
$x\equiv\ell_{\rm opt}/r^{d_{\rm opt}}$, we show in Fig.~\ref{Pi}
$\Pi(x,\lambda,A)$ for $d=2$ and the fit lines for the exponents of
both $f_1$ and $f_2$, which are reported in Table~\ref{table1}.  The
values of $\phi_{\rm opt}$ we calculate are close to the values predicted
by Eq.~(\ref{phi_formula}) for $d=2$,~3.

\section{Comparison between Flow in Percolation and the Optimal path}  
\label{comparison}

We now study the PDF $P(\ell_{\rm tr}|r,L)$ with the purpose of comparing it
to $P(\ell_{\rm opt}|r,L)$, and address analyze the
detailed conditions under which optimization and flow in percolation
occur.  Our analysis (see Sec.~\ref{discussion}) explains the 
differences we observe, and also the ``right way'' in which the two problems 
become equivalent.

Since simulations for flow on percolation clusters are performed, we
describe the two-dimensional case of the algorithm~\cite{Lopez}.  
We represent the reservoir
as a two-dimensional site percolation cluster, and choose sites at
$(-r/2,0)$ and $(r/2,0)$, denoted by $A$ and $B$, respectively, to be
the injection and extraction well positions.  Points $A$ and $B$ are
separated by a geometric distance $r$, and the system box has corners at
$(\pm L/2,\pm L/2)$.  We construct percolation clusters at $p_c$ using
the Leath algorithm \cite{Leath}.

To model tracer motion we use the analogy with electrical circuits,
where for each bond, the pressure drop corresponds to the voltage
difference, and the flow corresponds to the electrical current on the
bond.  A pressure difference between sites $A$ and $B$ drives the
tracer.  For each realization, $10^{4}$ tracers are introduced at site
$A$, and then collected at site $B$.  The set of all sites through which
there is a non-zero current defines the cluster backbone of $M_B$ sites.

The ``pressure'' difference across bonds is equivalent to a ``voltage''
difference, so by solving Kirchhoff's equations on the backbone, we
obtain the potential (pressure) drops $\Delta V$ over all bonds for a
given realization. Due to the nature of the model, no turbulence or other complex
fluid flow effects are considered, which is equivalent to assuming
laminar flow inside the system.
Additionally, given that the bonds have vanishing radius, the
tracer flow ``perfectly mixes'' at the nodes.
For site $i$ having $s_i$ outgoing bonds, the tracer
selects a bond with a probability
\begin{equation}
w_{ij}\equiv\frac{\Delta V_{ij}}{\sum_{j}\Delta V_{ij}} \hspace{0.5cm} 
[j=1,\dots,s_i;~i=1,\dots,M_B].
\label{probflow}
\end{equation}
For incoming bonds, $w_{ij}=0$. This guaranties that the tracer dynamics 
are completely convective, i.e., with infinite Peclet number~\cite{peclet,difussion}.
The total traveling length of a tracer is the number of bonds of the path 
connecting $A$ and $B$, chosen by this tracer.  
Since the particles do not interact with one another, it
is equivalent to launching one particle at a time into the cluster.
This procedure is known as \emph{particle launching algorithm}
\cite{Koplik,Sahimi_pla}.  We determine the probability distribution of
the tracer traveling lengths $P(\ell_{\rm tr}|r,L)$ by counting the number of
particles that travel from site $A$ to site $B$ along a path of length
$\ell_{\rm tr}$, over all the particles and all realizations of the
percolation cluster.

\begin{figure}
\epsfig{file=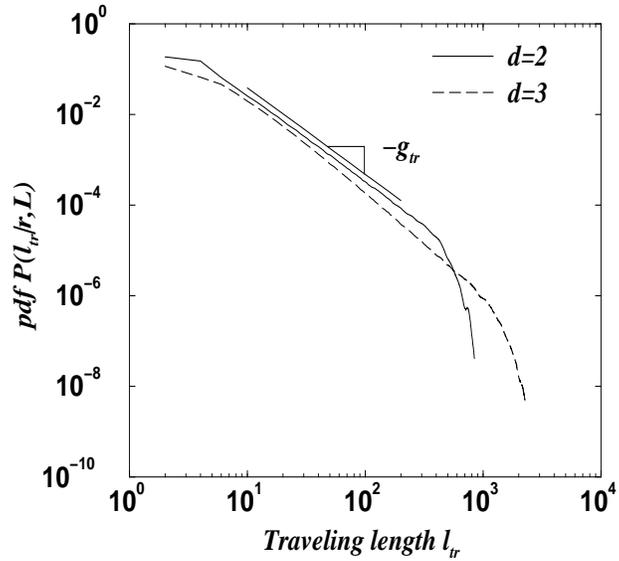,height=7.5cm,width=8.1cm}
\caption{Probability distribution $P(\ell_{\rm tr}|r,L)$ for $r=2$ and $L=128$
  in $d=2$ and $d=3$. In a similar fashion as for
  $P(\ell_{\rm opt}|r,L)$ we see a power law region that we characterize
  by exponent $g_{\rm tr}$.  Another feature of this plot is the increasing
  steepness of $P(\ell_{\rm tr}|r,L)$ as $d$ increases (implying $g_{\rm tr}$
  increases with $d$), a feature for which $P(\ell_{\rm opt}|r,L)$ has
  the opposite behavior, as $g_{\rm opt}$ decreases with $d$.}
\label{Pl_t}
\end{figure}

The form of $P(\ell_{\rm tr}|r,L)$ for the two-dimensional case was suggested
in \cite{Lee}. Here, we extend these results to $d=3$
(Fig.~\ref{Pl_t}).  Once again, the power law and stretched exponential
behaviors are present. The scaling of the most probable tracer path
length is given by $\ell^*_{\rm tr}\sim r^{d_{\rm tr}}$.  These results yield
\begin{equation}
P(\ell_{\rm tr}|r,L)\sim\frac{1}{r^{d_{\rm tr}}}
\left(\frac{\ell_{\rm tr}}{r^{d_{\rm tr}}}\right)^{-g_{\rm tr}}
h_1\left(\frac{\ell_{\rm tr}}{r^{d_{\rm tr}}}\right)
h_2\left(\frac{\ell_{\rm tr}}{L^{d_{\rm tr}}}\right),
\label{ansatz_l_t}
\end{equation}
where functions $h_1$ and $h_2$ have the forms
\begin{equation}
h_1(z)=\exp(-\mu z^{-\phi_{\rm tr}}) \qquad \left[z\equiv\frac{\ell_{\rm
      tr}}{r^{d_{\rm tr}}}\right],
\end{equation}
and
\begin{equation}
h_2(u)=\exp(-\rho u^{\psi_{\rm tr}}) \qquad \left[u\equiv\frac{\ell_{\rm
      tr}}{L^{d_{\rm tr}}}\right].
\end{equation}
Arguments similar to those leading to Eq.~(\ref{Pi_function}) indicate
how to determine the exponents $\phi_{\rm tr}$ and $\psi_{\rm tr}$,
reported in Table~\ref{table1}.

\begin{figure}
\epsfig{file=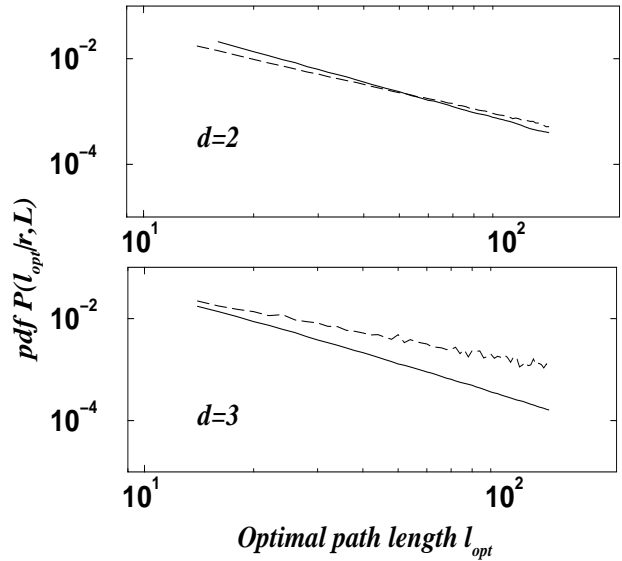,height=7.5cm,width=8.1cm}
\caption{
The power law tails of $P(\ell_{\rm tr}|r,L)$ (solid) and
  $P(\ell_{\rm opt}|r,L)$ (dashed) in $d=2$ and 3. The upper pair is
  for $d=2$ with $L=256$ and $r=4$.  The lower pair is for $d=3$ with
  $L=128$ and $r=2$. 
}
\label{Pl_opt_l_t}
\end{figure}

The power law region is characterized by the exponent $g_{\rm tr}$,
which is different from $g_{\rm opt}$ (see Table~\ref{table1}).  We
present in Fig.~\ref{Pl_opt_l_t} curves for the power law regime for
both $P(\ell_{\rm tr}|r,L)$ and $P(\ell_{\rm opt}|r,L)$ in $d=2$ and 3.
In Table \ref{table1} we see the difference in the slope of the power law decay
between $P(\ell_{\rm tr}|r,L)$ and $P(\ell_{\rm opt}|r,L)$.  Moreover,
as $d$ increases, $g_{\rm opt}$ decreases and $g_{\rm tr}$ increases,
indicating the differing behaviors for the two problems.  In the next
section, we explain the origin of the differences, how these differences
can be removed, and under which conditions the two problems coincide.

\section{Discussion}
\label{discussion}

The numerical results presented above show the difference in the values of the
scaling exponents $g_{\rm opt}$ and $g_{\rm tr}$ of the
distributions. To understand these differences, we now elaborate on the
characteristics of the optimal path problem in comparison to those of
tracer paths in percolation.

\begin{figure}
\epsfig{file=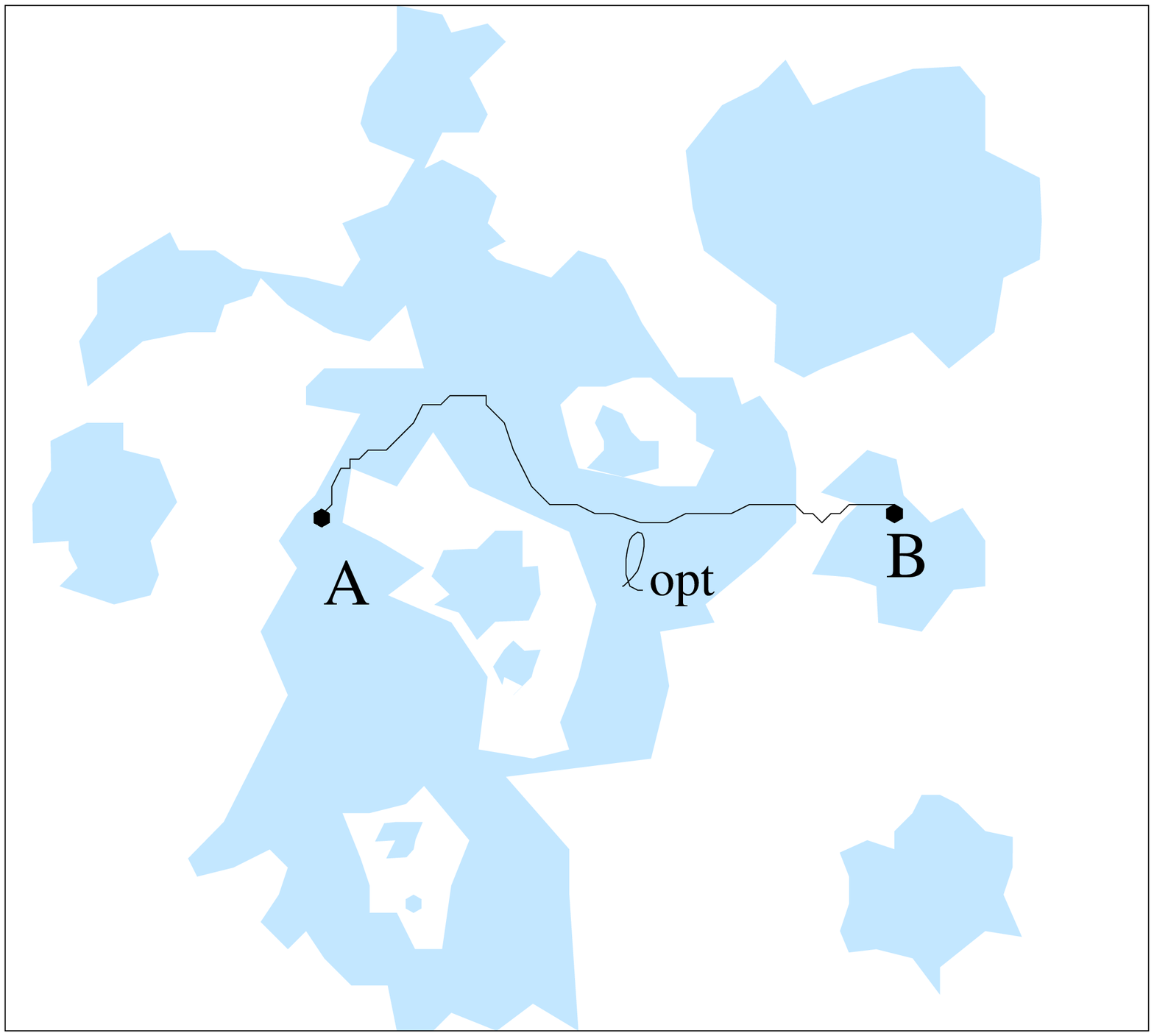,height=7.5cm,width=8.1cm}
\epsfig{file=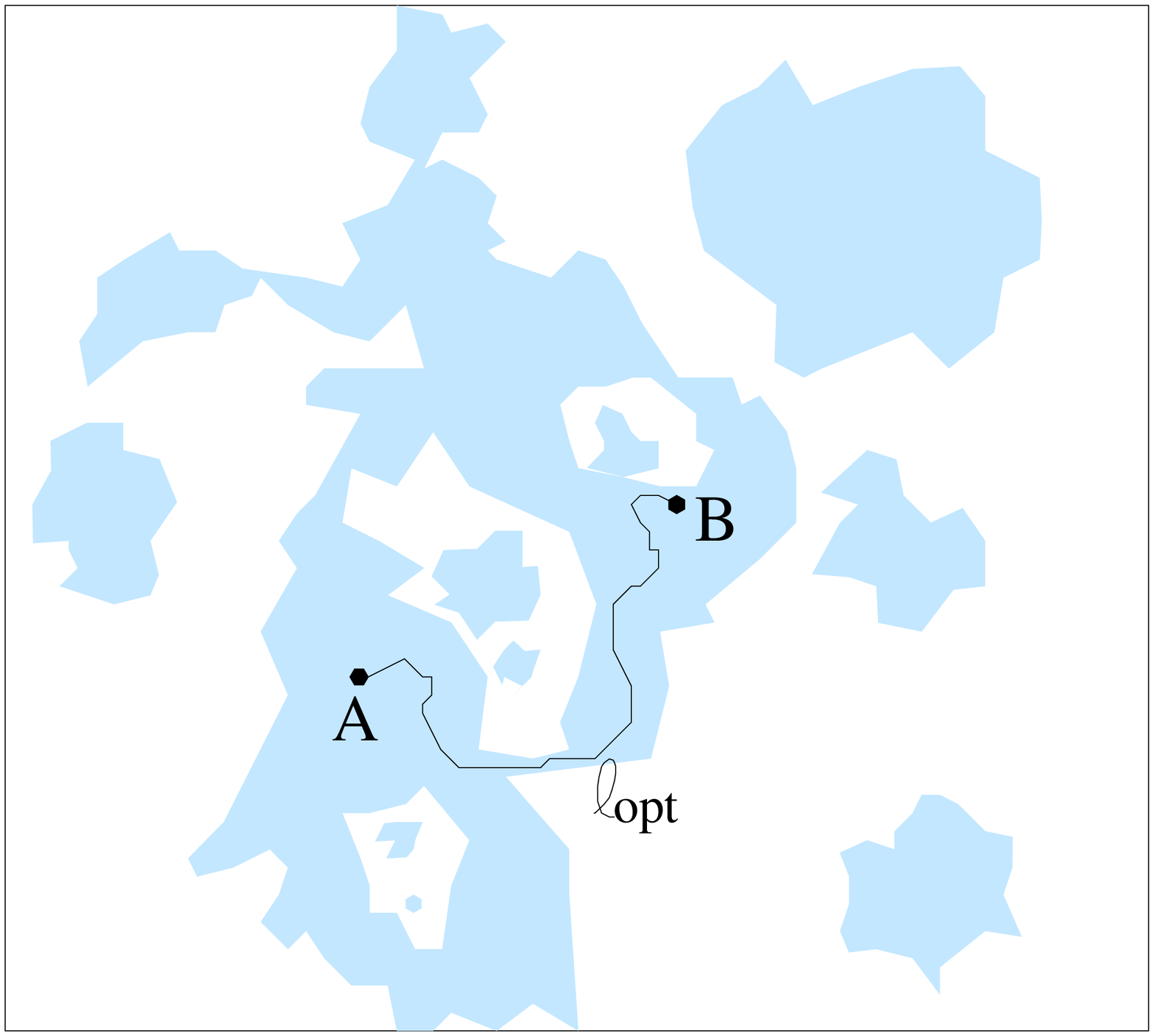,height=7.5cm,width=8.1cm}
\caption{(a) Schematic of occupied sites for $p$ below the percolation
  threshold $p_c$, and the optimal path in strong disorder.
  The darker regions represent sites that are still
  present when $p_c$ is reached. We see in this case that $\ell_{\rm
    opt}$ must cross the region above $p_c$ (i.e., leave the cluster) to
  connect $A$ and $B$.  (b) If sites $A$ and $B$ are chosen within the a
  cluster below $p_c$, the optimal path does not leave the cluster
  because that would increase the cost.  }
\label{schematic}
\end{figure}

In Fig.~\ref{schematic}(a) we represent the optimal path in strong
disorder, where the dark areas represent regions with site weights
$\epsilon_i= e^{ax_i}$ with $x_i\le p_c$, and the white areas regions with
site weigths $\epsilon=e^{ax_i}$ with $x_i>p_c$. 
Typically, the arbitrary choice of $A$ and $B$ may
lead to a path connecting them that requires visiting regions with site
weights $\epsilon>e^{ap_c}$.  In contrast, the tracers inside
percolation clusters must, by definition, travel on the {\it same\/}
percolation cluster (spanning or otherwise), because the flow takes
place only if there is a percolating path between $A$ and $B$.
Therefore, this difference between the flow and optimal path problems
presents a possible explanation for the differences between $P(\ell_{\rm
opt}|r,L)$ and $P(\ell_{\rm tr}|r,L)$.  Optimal paths tend to be longer
because they are able to visit more sites of the lattice and are
therefore of longer length, whereas tracers in percolation flow are
confined to a given cluster, and their traveling lengths are much more
limited. These features intuitively explain why $g_{\rm tr}$ is larger
than $g_{\rm opt}$.

\begin{figure}
\epsfig{file=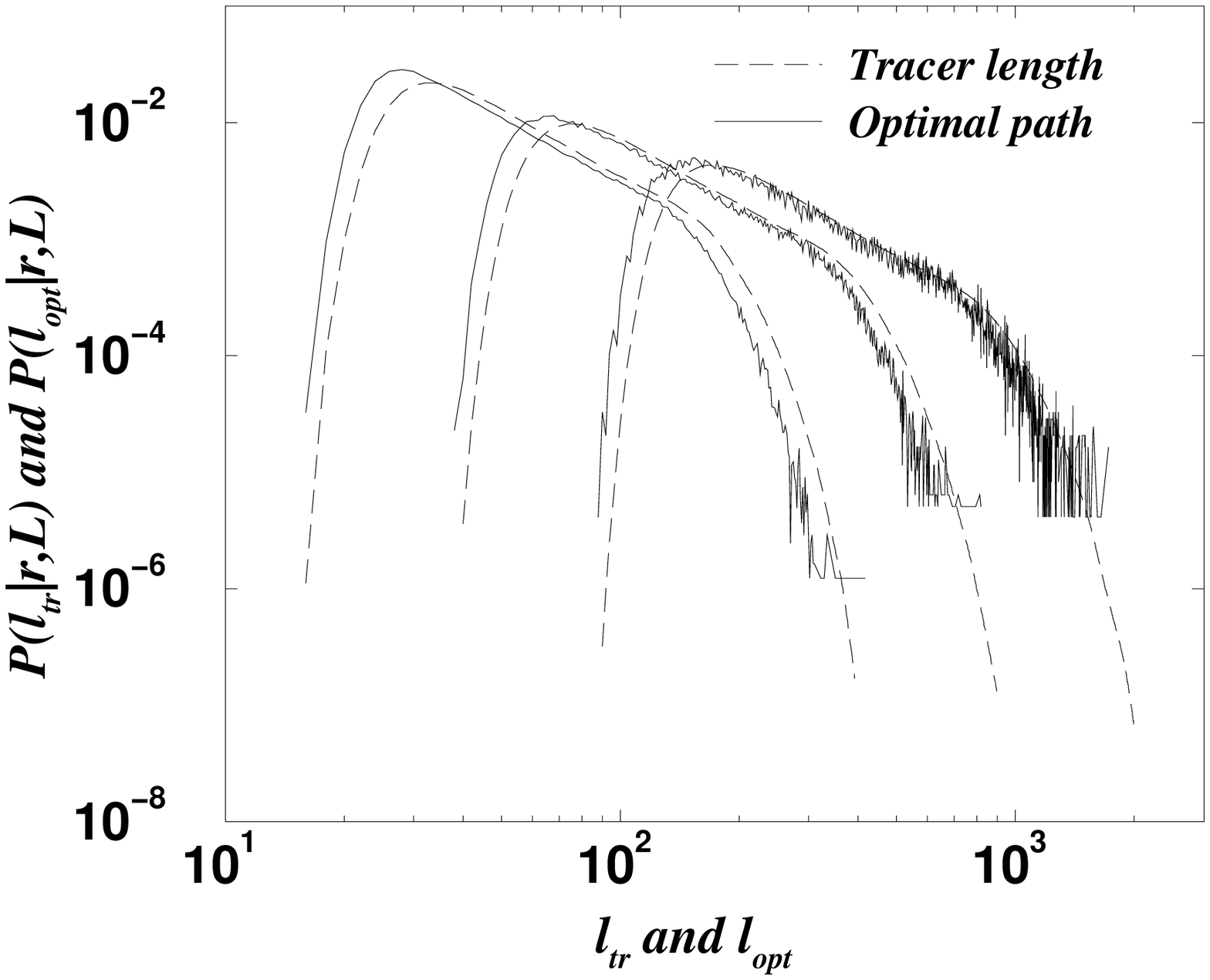,height=7.5cm,width=8.1cm}
\caption{Comparison of $P(\ell_{opt}|r,L)$ inside percolation 
with $P(\ell_{\rm tr}|r,L)$ for $(r=16, L=64)$, $(r=32,L=128)$ 
and $(r=64,L=256)$. 
The solid lines represent the optimal path distribution, and the
long dashed lines the tracer length distributions.
The values of $r$ and $L$ have a fixed ratio equal to $L/r=4$.
The similarity between distributions is clear, supporting our
hypothesis.
The small separation along the horizontal axis between $P(\ell_{opt}|r,L)$
and $P(\ell_{\rm tr}|r,L)$ (consistent for the three pairs of curves)
is due to non-universal details of the
two models. 
}
\label{Pl_Pltr_perc}
\end{figure}

The above considerations lead to the following hypothesis: if the
optimal path search is constrained to pairs of sites within regions of
the lattice that are part of the same cluster [Fig.~\ref{schematic}(b)],
then the scaling of $P(\ell_{\rm opt}|r,L)$ and $P(\ell_{\rm tr}|r,L)$
would coincide. To test
this, we present $P(\ell_{\rm opt}|r,L)$ and $P(\ell_{\rm tr}|r,L)$
in Fig.~\ref{Pl_Pltr_perc}, where the optimal paths satisfy the condition
that their highest weight is at or below percolation. This condition
forces the optimal paths to be inside percolation clusters. Indeed, for
this case (Fig.~\ref{Pl_Pltr_perc}), the two
quantities exhibit very similar behavior, supporting our
hypothesis. The exponent $g_{\rm opt}$ inside percolation now becomes
very close to $g_{\rm tr}$.  On the other hand, $d_{opt}$ does not
change, confirming the equivalence of the two problems.  We also have
similar results for three-dimensional lattices.

In summary, we have shown that $P(\ell_{\rm opt}|r,L)$ has a power law
tail with an exponent $g_{\rm opt}$ which decays as $d$ grows
and is different from the power law tail of
$P(\ell_{\rm tr}|r,L)$.
This difference seems to be related to the fact that the optimal path
crosses percolation clusters and thus tends to have longer lengths
compared with tracers which are always inside percolation
clusters. When $\ell_{\rm
opt}$ is measured only inside percolation clusters, 
our results suggest $P(\ell_{\rm
opt}|r,L)$ and $P(\ell_{\rm tr}|r,L)$ are equivalent and the two
problems possibly belong to the same universality class.

\newpage
\begin{table}
  \caption{
    Exponents characterizing $P(\ell_{\rm opt}|r,L)$ and $P(\ell_{\rm tr}|r,L)$,
    which are defined in the text.
    The value of
    $g_{\rm opt}$ is determined from $P(\ell_{\rm 
      opt}|r=4,L=256)$ in Fig.~\protect\ref{Pl_l_s4}(a), for which the power 
    law region is the longest. The values of $\phi_{\rm opt}$ and $\psi_{\rm
      opt}$ are from Fig.~2.}
\begin{tabular}{|c||l|l|l|l|l|}
\hline
\multicolumn{6}{|c|}{\textbf{Optimal Path in Strong Disorder 
(``Static'')}}\\ \hline
$d$&$d_{\rm opt}$&$g_{\rm opt}$&$\phi_{\rm opt}$ (calculated)
&$\phi_{\rm opt}=\frac{1}{d_{\rm opt}-1}$&$\psi_{\rm opt}$ (calculated)
\\ \hline
2&$1.22\pm 0.01$\cite{Cieplak,Buldyrev}&$1.55\pm 0.05$ 
&$4.8\pm 0.5$&$4.55\pm0.21$&$5.3\pm0.3$ \\ \hline
3&$1.42\pm 0.02$\cite{Cieplak2}&$1.37\pm 0.05$
&$2.1\pm0.1$&$2.3\pm0.1$&$4.3\pm0.3$ \\ \hline
\multicolumn{6}{|c|}{\textbf{Optimal Path in Strong Disorder inside 
Percolation (``Modified Static'')}}\\ \hline
2&$1.21\pm 0.02$&$1.82\pm0.05$&$4.9\pm0.4$&$4.76\pm0.45$&$2.3\pm0.4$ \\ \hline
3&$1.40\pm 0.03$&$2.2\pm 0.1$&
$2.0\pm 0.1$&$2.5\pm 0.2$&$3.6\pm 0.2$ \\ \hline
\multicolumn{6}{|c|}{\textbf{Tracer Path (``Dynamic'')}}\\ \hline
$d$&$d_{\rm tr}$&$g_{\rm tr}$&$\phi_{\rm tr}$ (calculated)&
$\phi_{\rm tr}=\frac{1}{d_{\rm tr}-1}$&$\psi_{\rm tr}$ (calculated)\\ \hline
2&$1.21\pm 0.02$\cite{Lee}&$1.82\pm0.05$ 
&$4.7\pm0.4$&$4.76\pm0.45$&$2.7\pm0.2$ \\ \hline
3&$1.37\pm 0.05$&$2.23\pm 0.09$
&$1.81\pm 0.02$&$2.7\pm0.4$&$3.46\pm0.04$ \\ \hline
\end{tabular}
\label{table1}
\end{table}

\clearpage

\clearpage

\clearpage

\clearpage

\clearpage

\clearpage


\begin{thebibliography}{99}
\bibitem{Text1} Since most of the computational demands come from
checking connectivity between $A$ and $B$, we use the method in
Ref.~\protect\cite{Buldyrev-algorithm}, which consists of finding the
shortest path between $A$ and $B$, and then, as long as none of the
sites of this shortest path are bombed, it is not necessary to check
connectivity.  Only when a site of the shortest path is bombed,
connectivity is checked. If there is a new shortest path, the site is
eliminated, but if there is no alternative path, then the site belongs
to the optimal path and the shortest path is kept.
\bibitem{peclet} The P\'{e}clet number is defined as the ratio between 
the characteristic length scale times the average fluid velocity, divided by the 
molecular diffusion coefficient.
Infinite P\'{e}clet number is equivalent to saying that no diffusion 
occurs in the system, and the problems is completely convective.
\bibitem{difussion} The difusive regime, of great relevance in many systems,
does not exhibit any relation to the connection between the convective 
tracer flow and the optimal path problem that we address.
\end{thebibliography}
\end{document}